\documentclass{article}
\usepackage[utf8]{inputenc}
\usepackage[T1]{fontenc}
\usepackage[T1]{fontenc}
\usepackage{graphicx}
\usepackage{amsmath,amssymb}
\usepackage{rotating}
\usepackage{color}
\usepackage[round]{natbib}
\usepackage{rotating}
\usepackage{multirow}
\usepackage[table]{xcolor}
\usepackage{caption}
\usepackage{chngcntr}
\usepackage{bm}
\usepackage{courier}
\usepackage{cite}
\usepackage{bm}
\usepackage{verbatim}
\usepackage[top=1in, bottom=1in, left=1in, right=1in]{geometry}
\usepackage[doublespacing]{setspace}
\usepackage[pdftex]{hyperref}
\usepackage{algorithm}
\usepackage{algpseudocode}

\begin{document}

\begin{center}
\begin{LARGE}
\begin{spacing}{1.5}
\textbf{The Unconditional Performance of Control Charts for Zero-Inflated Processes with Estimated Parameters}
\end{spacing}
\end{LARGE}
\end{center}

\begin{center}
$^{1}$Athanasios C. RAKITZIS \footnote{Corresponding author. E-mail address: \href{mailto:arakitz@unipi.gr}{arakitz@unipi.gr}. \\ This is an original manuscript of an article that will be published by Taylor \& Francis in \textit{Statistical Methods and Applications in Systems Assurance \& Quality} of the Book Series \textit{Advanced Research in Reliability and System Assurance}}, $^{2}$Eftychia MAMZERIDOU, $^{3}$Petros E. MARAVELAKIS  
\end{center}

\begin{center}
$^{1}$Department of Statistics and Insurance Science, University of Piraeus, Piraeus 18534, Greece. \\
$^{2}$Department of Statistics and Actuarial-Financial Mathematics, University of the Aegean, Karlovasi 83200, Samos, Greece. \\
$^{3}$Department of Business Administration, University of Piraeus, Piraeus 18534, Greece. \\
\end{center}

\begin{abstract}
Control charts for zero-inflated processes have attracted the interest of the researchers in the recent years. In this work we investigate the performance of Shewhart-type charts for zero-inflated Poisson and zero-inflated Binomial processes, in the case of estimated parameters. This is a case that usually occurs in practice, especially prior to starting the process monitoring. Using Monte Carlo simulation we evaluate charts' performance under an unconditional perspective and provide guidelines for their use in practice. We examine both the in-control and the out-of-control performance.
\end{abstract}


\section{Introduction}
\label{sec:Intro}
\textcolor{black}{Control charts are} the main tool of Statistical Process Control\index{statistical process control} (SPC) and can be used for monitoring a process in order to detect changes in it. Walter A. Shewhart \textcolor{black}{(\citet{shewhart1926quality})} was the first who proposed the use of control charts, back in the 1920's. The most common and \textcolor{black}{often used} are the Shewhart-type charts because of their ease in design and interpretation. Two main categories of control charts are the variables charts\index{control charts!for variables} and the attributes charts\index{control charts!for attributes}. Control charts for variables are used when the quality characteristic $X$ (random variable, r.v.) follows a continuous distribution while if $X$ follows a discrete distribution then the respective chart is a control chart for attributes. Control charts for attributes are used for process monitoring when the data collected by the process are counts. For example, attributes control charts are used for monitoring the number of non-conforming items in a sample, the number of non-conformities in each inspected unit, the number of new daily infections in \textcolor{black}{a} specific region or the daily number of logins in the computers of a workstation. 

The $np$- and $c$-chart are two common Shewhart-type attributes charts\index{control charts!$np$-chart}\index{{control charts!$c$-chart}}. The first is used when $X$ is a binomial random variable (such as the number of non-conforming items in \textcolor{black}{the sample}) while the second is used when $X$ is a Poisson random variable (such as the number of emergency calls in a hospital during the night shift). See \citet{montgomery2009} for the basic properties of the $np$- and $c$-chart. 

However, there are several processes where the $np$- and $c$-chart cannot be used for their monitoring because the distributional assumption of the binomial and the Poisson distribution is violated. For example, when there is overdispersion in the data or if there is an excessive number of zero values (or, sometimes, both). Therefore, \textcolor{black}{practitioners} have to select first the model that fits better their data and then develop a control chart based on this model. 

Zero-inflated models\index{zero-inflated models} (see \citet{johnson2005}, pp. 351-356) have been recommended as alternative models that take into account the excessive number of zeros as well as the overdispersion that \textcolor{black}{is caused} by them. This excess in zeros is attributed to the technological progress and automation of manufacturing technology which led to high-quality (or high-yield) processes\index{high-quality processes}; processes with a very low probability of \textcolor{black}{non-conforming items} or with a very low rate in the occurrence of non-conforming items. A zero value occurs when there are no non-conforming items in a sample or no non-conformities in the inspected unit.

\textcolor{black}{In recent years, there has been an increased interest in} control charts for monitoring zero-inflated processes\index{zero-inflated processes}. See the recent overview by \citet{mahmood2019models}. However, the majority of these works is about control charts for Phase II monitoring\index{Phase II}, where the basic assumption is the values of the process parameters are known. This is a strong assumption which rarely (if ever) is met in practice. According to \citet{montgomery2009} \textcolor{black}{(see also \citet{jensen2006})} there is significant difference between the theoretical and the actual performance of a control chart, \textcolor{black}{which is attributed to the presence of estimation error}. With the term theoretical we refer to the case where process parameters are known (or Case K\index{case K}) while the term actual performance refers to the case where the process parameters are unknown and must be estimated (Case U\index{case U}). \textcolor{black}{Albers et al.~\citet{albers2004parametric} discriminated the estimation error\index{estimation error} into two distinct sources, namely \textit{model error}\index{estimation error!model error} and \textit{stochastic error}\index{estimation error!stochastic error}. The first is caused by incorrect model assumptions (e.g. the true distributional model is much different than the selected one) while the latter is due to estimation uncertainty and is related to the selected Phase I sample. In this work the focus is solely on the stochastic (estimation) error}. He et al.~\citet{he2003} where the first who studied the estimation effect\index{estimation effect} on the performance of a Shewhart-type control chart for monitoring a zero-inflated Poisson process. \citet{rakitzis2016effect} extended the work of \citet{he2003} and studied also the performance of a Shewhart-type chart for monitoring a zero-inflated binomial process. Both works assume that only one of the two parameters of these zero-inflated processes is unknown. Recently, \citet{Mamzeridou2023zipU} investigated the performance of an upper one-sided Shewhart chart for monitoring a zero-inflated Poisson process, in the case where both process parameters are unknown.  

In this work we consider two popular zero-inflated models (namely, the zero-inflated Poisson and the zero-inflated binomial distribution) and study the statistical design and performance of the corresponding Shewhart-type charts in Case U. It should be mentioned that control charts for monitoring a zero-inflated binomial process in Case U \textcolor{black}{have not been considered so far}. In addition, we consider two estimation methods, maximum likelihood estimation (MLE) and method of moments (MoM), while the proposed chart is two-sided with Shewhart-type limits, instead of probability limits. The main difference with the work of \citet{Mamzeridou2023zipU} is that in this work we evaluate the performance of the charts in Case U \textcolor{black}{from a different perspective}. Thus, different performance measures are used. This will be explained later.

The structure of this paper is: In Section \ref{sec:CaseK} we provide, in brief, the basic properties of the Shewhart charts for monitoring a zero-inflated Poisson (Section \ref{sec:CaseKZIP}) and a zero-inflated binomial (Section \ref{sec:CaseKZIB}) process, in the case of known parameters (Case K), along with the appropriate measures for their performance (Section \ref{sec:CaseKRL}). In Section \ref{sec:CaseU}, we describe the implementation of the charts in Case U, for each of the considered processes, providing also the estimators under each case. Using these estimates we derive the basic properties of the considered Shewhart charts in the case of estimated parameters. Also, we present the steps for designing the charts as well as the appropriate performance measures, under an unconditional perspective. In Section \ref{sec:numerics} we present the findings of a simulation study regarding the in-control (IC)\index{in-control (IC)} and the out-of-control (OOC)\index{out-of-control (OOC)} performance of the considered charts in Case U. The results in the case of monitoring a ZIP process are given in Section \ref{sec:numericsZIP} while in Section \ref{sec:numericsZIB} we provide the results in the case of a ZIB process. A brief discussion on the OOC performance of the charts is given in Section \ref{sec:ooc} while conclusions are summarized in Section \ref{sec:conclusions}.

\section{Shewhart Charts for Monitoring ZIP and ZIB Processes} \label{sec:CaseK} 

\subsection{The Shewhart Chart for ZIP Processes (Case K)}\label{sec:CaseKZIP}
The zero-inflated Poisson (ZIP)\index{zero-inflated Poisson} distribution is a generalization of the Poisson distribution\index{Poisson distribution} that can be used to model count data containing an excessive number of zeros. By definition, if $X$ is a ZIP random variable, it is defined on $\left\{0,1,\ldots\right\}$ (as for the standard Poisson distribution) and its probability mass function (p.m.f.)\index{probability mass function} is given by 
\begin{equation}
	f_{ZIP}\left(x\right.\left|\phi,\lambda\right)=
	\left\{\begin{array}{rcl}
	\phi + \left(1-\phi\right) f_{P}\left(0\right.\left|\lambda\right), & x = &  0 \\
	\left(1-\phi\right) f_{P}\left(x\right.\left|\lambda\right), & x = & 1,2,...
	\end{array}\right.,
\end{equation}
%
where $f_{P}\left(x\right.\left|\lambda\right)$ is the p.m.f. of the standard Poisson distribution with parameter $\lambda >0$ and $\phi \in \left[0,1\right]$ is the zero-inflation parameter\index{zero-inflation parameter}. If $\phi =0$, then the ZIP distribution coincides with the standard Poisson distribution. If $\phi =1$, then the ZIP distribution reduces to the Dirac distribution on $x=0$. Moreover, it is not difficult to verify that the cumulative distribution function (c.d.f.)\index{cumulative distribution function} of $X$ is given by:
\begin{equation}
F_{ZIP}\left(x\right.\left|\phi,\lambda\right)=\phi +(1-\phi)F_{P}\left(x\right.\left|\lambda\right),
\end{equation}
where $F_{P}\left(x\right.\left|\lambda\right)$ is the c.d.f. of the Poisson distribution with parameter $\lambda$. The mean\index{zero-inflated Poisson!mean} and the variance\index{zero-inflated Poisson!variance} of the ZIP distribution with parameters $\left(\phi,\lambda\right)$ are, respectively, given by
\begin{equation}
E(X)=\lambda\left(1-\phi\right),\;\;V(X)=\lambda\left(1+\lambda\phi\right)\left(1-\phi\right).
\end{equation}

 Let us now assume that we want to set-up a Shewhart control chart\index{Shewhart control chart} for monitoring a ZIP process. In the case of known parameters (Case K), the upper and the lower control limits\index{control limits!upper control limit ($UCL$)}\index{control limits!lower control limit ($LCL$)} (denoted as $UCL_{ZIP}$, $LCL_{ZIP}$, respectively) are given by \textcolor{black}{(see, for example, \citet{rakitzis2016effect}, page 4196):}
\begin{equation}
\label{eq:UCLzip}
UCL_{ZIP}=\left\lfloor \lambda_{0}(1-\phi_0)+L\sqrt{\lambda_{0}(1+\lambda_{0}\phi_0)(1-\phi_0)}\right\rfloor,
\end{equation}
\begin{equation}
\label{eq:LCLzip}
LCL_{ZIP}=\max\left(0, 
\left\lceil\lambda_{0}(1-\phi_0)-L\sqrt{\lambda_{0}(1+\lambda_{0}\phi_0)(1-\phi_0)}\right\rceil\right),
\end{equation}
where $\left\lfloor \ldots\right\rfloor$ and $\left\lceil \ldots\right\rceil$ denote the rounded down and rounded up integer, $\phi_0$ $\lambda_{0}$ are the in-control values of the process parameters $\phi$ and $\lambda$, while $L>0$ is a constant that plays the role of chart's design parameter. These limits are also known as \textit{Shewhart-type} or $L\sigma$ limits\index{control limits!Shewhart-type}. It should be noted that in \citet{Mamzeridou2023zipU} the authors used only one probability limit and not Shewhart-type limits. \textcolor{black}{Also, given $\phi_0$ and $\lambda_0$, the value of $L$ determines the limits of the interval $[LCL_{ZIP},UCL_{ZIP}]$. Thus, it is directly related to the false alarm rate (FAR) of the chart. Recall that the FAR is the probability for the chart to give an OOC signal when the process is actually IC. The largest the $L$ the smallest the FAR}.

Let $Y_{1},Y_{2},\ldots$ be independent random variables such that $Y_{i}\sim ZIP(\phi_1,\lambda_{1})$, i.e. a ZIP distribution with parameters $(\phi_1,\lambda_{1})$, where, in general, $\phi_1\neq\phi_0$, $\lambda_{1}\neq\lambda_0$ and $\phi_1 \in [0,1]$, $\lambda_1>0$. If $\phi_1=\phi_0$ and $\lambda_1=\lambda_0$, then the process is IC\index{in-control (IC)}, otherwise it is OOC\index{in-control (IC)}. Then, the probability $\beta$ which is the probability that the number $Y_{i}$ of non-conformities is within the interval $[LCL_{ZIP},UCL_{ZIP}]$ equals
\begin{eqnarray} \nonumber
\label{eq:pOOCzip}
\beta & = & P\left(LCL_{ZIP}\leq Y_{i} \leq UCL_{ZIP}\right) \\
          & = & F_{ZIP}\left(\left.UCL_{ZIP}\right|\phi_1,\lambda_{1}\right)-F_{ZIP}\left(\left.LCL_{ZIP}-1\right|\phi_1,\lambda_{1}\right)
\end{eqnarray}

\subsection{The Shewhart Chart for ZIB Processes (Case K)}\label{sec:CaseKZIB}
The zero-inflated binomial (ZIB)\index{zero-inflated binomial} distribution is a generalization of the binomial distribution\index{binomial distribution} that takes into account the excessive number of zeros, compared to the usual binomial distribution. By definition, if $X$ is a ZIB random variable, it is defined on $\left\{0,1,\ldots,n\right\}$ (same as the binomial distribution) and its p.m.f.\index{probability mass function} is given by 
\begin{equation}
	f_{ZIB}\left(x\right.\left|\phi,n,p\right)=
	\left\{\begin{array}{rcl}
	\phi + \left(1-\phi\right) f_{B}\left(0\right.\left|n,p\right), & x = &  0 \\
	\left(1-\phi\right) f_{B}\left(x\right.\left|n,p\right), & x = & 1,2,...,n
	\end{array}\right.,
\end{equation}
\noindent where $f_{B}\left(x\right.\left|n,p\right)$ is the p.m.f. of the binomial distribution $B(n,p)$ with parameters $n\in\{1,2,\ldots\}$ and $p\in (0,1)$. Parameter $p$ is the success probability (i.e. the probability of an item being non-conforming) while $\phi$ is the zero-inflation parameter\index{zero-inflation parameter}. If $\phi =0$, the ZIB distribution coincides with the binomial distribution with parameters $n$ and $p$ while, for $\phi =1$, the ZIB distribution reduces to the Dirac distribution on $x=0$. The c.d.f. of $X$ is given by
\begin{equation}
F_{ZIB}\left(x\right.\left|\phi,n,p\right)=\phi +(1-\phi)F_{B}\left(x\right.\left|n,p\right),
\end{equation}
where $F_{B}\left(x\right.\left|n,p\right)$ is the c.d.f.\index{cumulative distribution function} of $B(n,p)$. The mean\index{zero-inflated binomial!mean} and the variance\index{zero-inflated binomial!variance} of the ZIB distribution with parameters $\left(\phi,n,p\right)$ are, respectively, given by the following two expressions
\begin{equation}
E(X)=np\left(1-\phi\right),\;\;V(X)=np\left(1-p+np\phi\right)\left(1-\phi\right).
\end{equation}

In a similar manner, the control limits\index{control limits!upper control limit ($UCL$)}\index{control limits!lower control limit ($LCL$)} of a the Shewhart chart for monitoring a ZIB process with known parameters are given by \textcolor{black}{(see, for example, \citet{rakitzis2016effect}, page 4196):}
\begin{equation}
\label{eq:UCLzib}
UCL_{ZIB}=\left\lfloor np_{0}(1-\phi_0)+L\sqrt{np_{0}(1-p_{0}+np_{0}\phi_0)(1-\phi_0)}\right\rfloor,
\end{equation}
\begin{equation}
\label{eq:LCLzib}
LCL_{ZIB}=\max\left(0, 
\left\lceil np_{0}(1-\phi_0)-L\sqrt{np_{0}(1-p_{0}+np_{0}\phi_0)(1-\phi_0)} \right\rceil\right),
\end{equation}
where $\phi_0$, $p_{0}$ are the in-control values of process parameters $\phi$ and $p$, while constant $L>0$ is the chart's design parameter. Let $Y_{1},Y_{2},\ldots$ be independent random variables with $Y_{i}\sim ZIB(\phi_1,n,p_{1})$, i.e., the number of non-conforming units in a sample of size $n$ is a ZIB random variable with parameters $(\phi_1,n,p_{1})$. Again, in general, $\phi_1\neq\phi_0$ and $p_1\neq p_0$ with $\phi_1 \in [0,1]$ and $p_1\in (0,1)$. If $\phi_1=\phi_0$ and $p_1=p_0$ the process is IC\index{in-control (IC)}, otherwise it is OOC\index{out-of-control (OOC)}. Then, the probability $\beta$ which is the probability that the number $Y_{i}$ of non-conforming items is within the interval $[LCL_{ZIB},UCL_{ZIB}]$ equals
\begin{eqnarray} \nonumber
\label{eq:pOOCzib}
\beta & = & P\left(LCL_{ZIB} \leq Y_{i} \leq UCL_{ZIB}\right) \\
          & = & F_{ZIB}\left(\left.UCL_{ZIP}\right|\phi_1,n,p_{1}\right)-F_{ZIP}\left(\left.LCL_{ZIP}-1\right|\phi_1,n,p_{1}\right)
\end{eqnarray}

\subsection{Run length properties (Case K)}\label{sec:CaseKRL}
The run length of the Shewhart charts for ZIP and ZIB processes with known parameters is a geometric random variable\index{geometric distribution} $N$ with parameter $1-\beta$, i.e., the p.m.f.\index{probability mass function} $f_N(\nu)$ and the c.d.f.\index{cumulative distribution function} $F_N(\nu)$ of $N$ are defined for $\nu =1,2,\ldots$ and they are, respectively, given by:
\begin{equation}
f_{N}(\nu)=(1-\beta)\beta^{\nu-1},\;\;F_{N}(\nu)=1-\beta^{\nu},
\end{equation}
\noindent where $\beta$ is the probability defined in equations (\ref{eq:pOOCzip}) and (\ref{eq:pOOCzib}). Therefore, the Average Run Length ($ARL$)\index{average run length} and the Standard Deviation Run Length ($SDRL$)\index{standard deviation run length} of the Shewhart charts for ZIP and ZIB processes with known parameters equal
\begin{equation}
E(N)=ARL=\frac{1}{1-\beta},\:\:\sqrt{V(N)}=SDRL=\frac{\sqrt{\beta}}{1-\beta}.
\end{equation}
Note also that the moment about the origin of order two of the run length distribution equals
\begin{equation}
\mu^{\prime}_2=E(N^2)=V(N)+E^2(N)=\frac{\beta}{(1-\beta)^2}+\frac{1}{(1-\beta)^2}=\frac{1+\beta}{(1-\beta)^2},
\end{equation}
\noindent where $\mu^{\prime}_k=E(N^k)$, $k=1,2,\ldots$. \textcolor{black}{It should be mentioned that since $\beta$ is given by equations (\ref{eq:pOOCzip}) or (\ref{eq:pOOCzib}) the run length distribution as well as the $ARL$ and $SDRL$ refer all to the OOC case. For $\phi_1=\phi_0$ and $\lambda_1=\lambda_0$ we get the respective IC measures.} 

\section{Shewhart Charts for Monitoring ZIP and ZIB Processes With Unknown Parameters} \label{sec:CaseU} 

In this section we discuss the properties of the two Shewhart charts, for ZIP and ZIB processes, in Case U. The two processes are treated individually. As already stated, for estimating the process parameters, we consider both the MLE method and the MoM.

\subsection{The Shewhart Chart for ZIP Processes (Case U)}\label{sec:CaseUZIP}
Suppose that we are interested in monitoring a ZIP process\index{zero-inflated Poisson} but the IC values of the process parameters are unknown. Thus, $\phi_0$ and $\lambda_0$ must be estimated from a Phase I\index{Phase I} sample. Let $X_1,X_2,\ldots,X_m$ be a random sample from $ZIP(\phi_0,\lambda_0)$, i.e., $X_i \sim ZIP(\phi_0,\lambda_0)$, $i=1,2,\ldots,m$. This is the Phase I sample which is used for estimating $\phi_0$ and $\lambda_0$. 

Using the MoM, the respective moment estimators (ME)\index{moment estimators!zero-inflated Poisson} for $\lambda_0$ and $\phi_0$ are given by:
\begin{equation}
\tilde{\lambda}=\frac{\overline{X^2}}{\bar{X}}-1,\;\tilde{\phi}=1-\bar{X}/\tilde{\lambda}_0,
\end{equation}
where 
\begin{equation}
\bar{X}=\frac{1}{m}\sum_{i=1}^{m}X_i,\;\;\overline{X^2}=\frac{1}{m}\sum_{i=1}^{m}X_i^2,
\end{equation}

Also, using the MLE method, the respective estimators\index{maximum likelihood estimators!zero-inflated Poisson} are not given in closed form but they are obtained numerically by solving the following non-linear system of equations:

\begin{equation}
\left\{
\begin{array}{l}
	\hat{\lambda}=\bar{X}^{+}(1-e^{-\hat{\lambda}})\\
\hat{\phi}=1-\bar{X}/\hat{\lambda}
\end{array}
\right.
\end{equation}
\noindent where $\bar{X}^{+}$ is the average of strictly positive $X$'s. 

When $\phi_{0}$, $\lambda_{0}$ are estimated by $\hat{\phi}$, $\hat{\lambda}$, the control limits of the Shewhart chart\index{Shewhart control chart} are denoted as $\widehat{LCL}_{ZIP}$, $\widehat{UCL}_{ZIP}$ and they are simply obtained by substituting $\phi_0$, $\lambda_0$ with $\hat{\phi}$ and $\hat{\lambda}$ in equations (\ref{eq:LCLzip}) and (\ref{eq:UCLzip}). In a similar manner, if $\phi_0$ and $\lambda_0$ are estimated by $\tilde{\phi}$, $\tilde{\lambda}$, the respective limits are denoted as $\widetilde{LCL}_{ZIP}$, $\widetilde{UCL}_{ZIP}$. It is not difficult to see that in Case U the control limits are based on the estimates of process parameters obtained from the available Phase I sample and thus, they are random variables. Given the estimates of $\phi_0$ and $\lambda_0$ it is possible to calculate the limits and proceed to Phase II process monitoring. 

It is not difficult to realize that different practitioners will end up with different estimates for the process parameters, because of the different Phase I samples. Thus, it is possible to use different control limits. In order to reflect the estimation error\index{estimation error} that it is attributed to the different estimates among all the practitioners, we will use the probability $\hat{\beta}=P(Y_i\in [\widehat{LCL}_{ZIP},\widehat{UCL}_{ZIP}]\left|\phi_1,\lambda_1\right.)$ (or $\tilde{\beta}=P(Y_i\in [\widetilde{LCL}_{ZIP},\widetilde{UCL}_{ZIP}]\left|\phi_1,\lambda_1\right.)$, in case of MoM estimators) to investigate the performance of the ZIP-Shewhart chart in Case U. Note also that this probability cannot be calculated in practice because $\phi_1$, $\lambda_1$ is unknown. 

The assessment of the control chart's performance in Case U is made under two perspectives, the `unconditional'\index{unconditional perspective} or the `conditional'\index{conditional perspective}. In Case U, the run length distribution of the ZIP-Shewhart chart is no longer a geometric one because the control limits are r.v. However, given the estimates $\hat{\phi}$, $\hat{\lambda}$ (or $\tilde{\phi}$, $\tilde{\lambda}$), the run length of the chart becomes \textit{conditional} on them and thus, it is known \textit{conditional run length}\index{conditional run length distribution}. Its distribution is now a geometric one. Using the conditional run length distribution it is possible to evaluate the performance of the chart under the `conditional' perspective. This approach was followed by \citet{Mamzeridou2023zipU}. See also \citet{zhao2016c,vakilian2018guaranteed} regarding the use of the conditional run length distribution to study the $c$-chart in Case U.

On the other hand, by averaging the conditional run length distribution over the (joint) distribution of $(\hat{\phi},\hat{\lambda})$ (or $(\tilde{\phi},\tilde{\lambda})$) we get the \textit{unconditional} run length distribution\index{unconditional run length distribution}. Using the unconditional run length distribution, the performance of the chart is evaluated under the unconditional perspective. This approach was followed by \citet{he2003} and \citet{rakitzis2016effect}. The majority of the literature on the performance of attributes control charts in Case U is based on the use of the `unconditional' run length distribution (see, for example, \citet{braun1999,testik2006effect,castagliola2012design,castagliola2014synthetic,wu2016run,johannssen2022performance} and references therein). An advantage of this perspective is that it can be applied before collecting the Phase I sample. Thus, it is possible to have practical guidelines regarding the size of the Phase I samples that gives the desired IC performance. For further details on the use of the unconditional run length distribution in studying the performance of control charts in Case U, see \citet{jensen2006}. Motivated by the previously mentioned works, we will use the unconditional (or marginal) run length distribution.   

In Algorithm \ref{alg:algo1} we provide the steps for the calculation of the appropriate performance measures of the ZIP-Shewhart chart in Case U. To the best of our knowledge, the joint distribution of $(\hat{\phi},\hat{\lambda})$ (or $(\tilde{\phi},\tilde{\lambda})$) is not known. For this reason, we use Monte Carlo simulation\index{Monte Carlo simulation}. The steps are given in the case of MLE method while in the case of the MoM, the steps are similar. In order to avoid additional complexities in notation we will use the notations $ARL$, $\mu^{\prime}_2$ and $SDRL$ in either Case K and Case U. However, in the latter case, these are the \textit{unconditional} measures\index{average run length!unconditional}\index{standard deviation run length!unconditional}. 

\begin{algorithm}[htbp]
\caption{Calculation of the unconditional performance measures - ZIP process}\label{alg:algo1}
\begin{algorithmic}[1]
\State Choose a Phase I sample of size $m$ from a $ZIP(\phi_0,\lambda_0)$ process.
\State Use the Phase I sample and estimate $\phi_0$, $\lambda_0$ by using the MLE method (and get the $\hat{\phi}$, $\hat{\lambda}$ estimates).
\State Given the values $L$, $\hat{\phi}$ and $\hat{\lambda}$, determine $\widehat{UCL}_{ZIP}$, $\widehat{LCL}_{ZIP}$ from equations (\ref{eq:UCLzip}) and (\ref{eq:LCLzip}), respectively.
\State Calculate the IC $ARL$ and the IC $\mu^{\prime}_2$ as $1/\left(1-\hat{\beta}\right)$ and $\left(1+\hat{\beta}\right)/\left(1-\hat{\beta}\right)^2$, respectively.
\State Repeat Steps 1-4 $T$ times and calculate (i) the \textit{unconditional} IC $ARL$ as the average of the $T$ values $1/\left(1-\hat{\beta}\right)$, (ii) the \textit{unconditional} IC $\mu^{\prime}_2$ as the average of the $T$ values $\left(1+\hat{\beta}\right)/\left(1-\hat{\beta}\right)^2$, (iii) the \textit{unconditional} IC $SDRL$ as $\sqrt{\mu^{\prime}_2-ARL^2}$, where $ARL$ and $\mu^{\prime}_2$ are those obtained in (i) and (ii).
\end{algorithmic}
\end{algorithm}

\subsection{The Shewhart Chart for ZIB Processes (Case U)}\label{sec:CaseUZIB}
Let $X_1,X_2,\ldots,X_m$ be a random sample from the zero-inflated binomial (ZIB)\index{zero-inflated binomial} distribution with in-control parameters $n$, $p_0$ and $\phi_0$, i.e., $X_i \sim ZIB(\phi_0,n,p_0)$, $i=1,2,\ldots,m$. Using the MoM\index{moment estimators!zero-inflated binomial}, the respective estimates of the IC process parameters are given by:
\begin{equation}
\tilde{p}=\frac{\overline{X^2}-\bar{X}}{(n-1)\bar{X}},\;\tilde{\phi}=1-\frac{(n-1)\bar{X}^2}{n\left(\overline{X^2}-\bar{X}\right)},
\end{equation}
where 
\begin{equation}
\bar{X}=\frac{1}{m}\sum_{i=1}^{m}X_i,\;\;\overline{X^2}=\frac{1}{m}\sum_{i=1}^{m}X_i^2,
\end{equation}

Also, similar to the case of ZIP process, the MLEs\index{maximum likelihood estimators!zero-inflated binomial} are not given in closed form but they are obtained numerically by solving the following non-linear system of equations:

\begin{equation}
\left\{
\begin{array}{l}
	\hat{p}=\bar{X}^{+}(1-(1-\hat{p})^n)\\
\hat{\phi}=1-\bar{X}/(n\hat{p})
\end{array}
\right.
\end{equation}
where $\bar{X}^{+}$ is the average of strictly positive $X$'s. When $\phi_{0}$, $p_{0}$ are estimated by $\hat{\phi}$, $\hat{p}$, the control limits of the Shewhart chart are denoted as $\widehat{LCL}_{ZIB}$, $\widehat{UCL}_{ZIB}$. Similarly, if the process parameters are estimated by $\tilde{\phi}$, $\tilde{p}$, the respective limits are denoted as $\widetilde{LCL}_{ZIB}$, $\widetilde{UCL}_{ZIB}$. Therefore, the control limits of the ZIB-Shewhart chart in Case U are calculated by simply replacing the (uknownn) values $\phi_0$ and $p_0$ with their respective estimates. Then, practitioners can proceed to the Phase II process monitoring.

The performance of the ZIB-Shewhart chart\index{Shewhart control chart} in Case U is evaluated in terms of the \textit{unconditional} $ARL$ and $SDRL$. The steps for their calculation are identical to those provided in Section \ref{sec:CaseUZIP} for the ZIP-Shewhart chart, after some straightforward but necessary modifications. Due to space economy, the details are left to the readers.

\section{Numerical Results} 
\label{sec:numerics} 

In this section, we present the results of an extensive simulation study, using 50000 simulation runs (i.e. $T=50000$), regarding the IC\index{in-control (IC)} performance of the ZIP- and ZIB-Shewhart charts in Case U. 

\subsection{The Unconditional Performance of the ZIP-Shewhart Chart}
\label{sec:numericsZIP}

In Tables \ref{tab:zipMLEu} and \ref{tab:zipMMu} we present the IC $ARL$ and $SDRL$ values for the ZIP-Shewhart chart\index{Shewhart control chart} in Case U, for $m \in \left\{100,200,500,1000,2000,5000\right\}$, $\lambda_{0} \in \left\{1,2,4,5,6,8\right\}$ and $\phi \in \left\{0.9,0.8,0.7\right\}$. The MLE method has been used to get the results in Table \ref{tab:zipMLEu} while the MoM has been used for the results in Table \ref{tab:zipMMu}. The first line in each cell gives the $ARL$ while the second line gives the $SDRL$. In the column `Case K' we provide the $ARL$ and $SDRL$ values in Case K while constant $L$, given in the respective column, has been determined in Case K (in two decimals accuracy) so as the IC $ARL$ is \textit{as close as possible} to the nominal value $ARL_0=370.4$, a value that appears very frequently in several SPC textbooks. It is well known (see, for example, \citet{rakitzis2016effect}) that due to the discrete nature of the ZIP distribution it is almost impossible to achieve \textit{exactly} the desired $ARL_0$ value. From the results in column 'Case K' in Table \ref{tab:zipMLEu} (and Table \ref{tab:zipMMu}) we deduce that there are cases where the IC $ARL$ in Case K is above 370.4 while for the remaining ones it is below.

It is not difficult to see that for the particular combinations $(\phi_0,\lambda_0)$, the IC $ARL$ values can be very different between the Case K and Case U. For example, from the results in Table \ref{tab:zipMLEu}, when $(\phi_0,\lambda_0)=(0.8,4)$ the IC $ARL$ ($SDRL$) in Case K is $ARL=234.04$ ($SDRL=233.54$) while when $m=200$ we have $ARL=566.39$ ($SDRL=1116.81$). Note also that for a relatively small Phase I sample (i.e. for $m=100$ or 200), the $ARL$ and $SDRL$ are very large, especially the latter. Actually, for large $\phi_0$ values (such as 0.9 or 0.8) and for Phase I samples of size $m=100$, the $SDRL$ values are extremely large. This is an indication of increased uncertainty regarding the actual IC performance of the chart. As $m$ increases, the IC $ARL$ values tend to decrease, but not in a monotonic way. This is common for attributes control charts with estimated parameters (see, for example, \citet{castagliola2012design}, \citet{rakitzis2016effect}, \citet{Mamzeridou2023zipU} and references, therein). Note also that as $m$ increases, it is not necessary that the IC $ARL$ and $SDRL$ obtained in Case U will converge to the respective IC $ARL$ and $SDRL$ values in Case K. Even for very large Phase I samples (such as for $m=2000$ or 5000), the $ARL$ and $SDRL$ can be significantly different than the respective values in Case K. See, for example, the case $(\phi_0,\lambda_0)=(0.8,1)$ or $(\phi_0,\lambda_0)=(0.7,2)$. This fact is also attributed to the discrete nature of the ZIP distribution.

\begin{table}[t!]
\tabcolsep6pt
\caption{IC $ARL$ and $SDRL$ values for the ZIP-Shewhart Chart in Case U when $L$ is from Case K - MLE method}
\begin{center}
\begin{tabular}{c|c|c|cccccc|c}\hline
						
$\phi_0$ &	$\lambda_0$ &	$L$ &	100 &	200 &	500 &	1000 &	2000 &	5000 &	Case K \\ \hline
0.9	&1	&6.66	&735.21	&426.62	&330.69	&323.88	&325.72	&326.11	&526.64 \\
	&	&	&3164.42	&1039.20	&502.21	&434.35	&431.99	&432.28	&526.14 \\
	&2	&6.41	&1268.37	&617.42	&435.46	&395.69	&392.04	&393.24	&189.92 \\
	&	&	&11282.42	&1739.44	&694.71	&518.63	&489.55	&489.74	&189.42 \\
	&4	&5.61	&1173.64	&568.04	&390.56	&348.66	&335.52	&335.86	&468.09 \\
	&	&	&14863.58	&1527.86	&614.13	&447.65	&392.86	&386.74	&467.59 \\
	&5	&5.72	&2775.20	&1043.62	&643.09	&555.63	&522.05	&511.23	&314.19 \\
	&	&	&32197.04	&3564.15	&1118.23	&752.59	&624.91	&589.38	&313.69 \\
	&6	&5.31	&1685.94	&714.62	&457.20	&396.67	&374.75	&366.50	&497.71 \\
	&	&	&14686.21	&2545.71	&764.17	&522.35	&439.39	&410.96	&497.21 \\
	&8	&5.15	&3201.76	&991.22	&575.55	&486.03	&450.12	&432.52	&292.56 \\
	&	&	&45373.05	&4210.97	&1025.86	&665.61	&538.16	&480.22	&292.06 \\ \hline
0.8	&1	&6.33	&1605.20	&1033.08	&811.91	&797.17	&801.98	&799.38	&263.32 \\
	&	&	&8203.23	&2641.74	&1245.20	&1115.59	&1118.12	&1116.19	&262.82 \\
	&2	&5.49	&1561.49	&943.59	&726.60	&700.33	&696.85	&692.86	&301.87 \\
	&	&	&7511.36	&2157.01	&1075.06	&911.12	&897.56	&894.40	&301.37 \\
	&4	&4.47	&850.31	&566.39	&449.27	&424.31	&420.75	&417.34	&234.04 \\
	&	&	&3305.00	&1116.81	&620.58	&518.11	&499.33	&496.14	&233.54 \\
	&5	&4.47	&1536.53	&927.90	&707.73	&650.36	&633.37	&634.67	&365.09 \\
	&	&	&6766.62	&2016.17	&1033.60	&808.04	&746.39	&744.54	&364.59 \\
	&6	&4.09	&875.03	&568.34	&446.35	&411.82	&403.97	&400.85	&248.86 \\
	&	&	&3104.88	&1129.01	&622.37	&498.47	&463.78	&458.92	&248.36 \\
	&8	&3.89	&1092.12	&667.40	&502.29	&459.47	&444.99	&440.59	&289.74 \\
	&	&	&4205.83	&1436.17	&717.51	&562.35	&505.80	&494.03	&289.24 \\ \hline
0.7	&1	&5.18	&785.25	&587.16	&535.49	&538.28	&537.27	&539.31	&175.55 \\
	&	&	&2352.71	&1179.19	&757.51	&474.96	&747.22	&748.72	&175.05 \\
	&2	&4.5	&693.97	&532.31	&465.54	&462.64	&468.50	&467.07	&201.24 \\
	&	&	&1799.96	&941.45	&615.07	&597.17	&601.32	&600.20	&200.74 \\
	&4	&3.66	&412.11	&328.80	&291.87	&287.96	&290.63	&294.41	&409.89 \\
	&	&	&896.25	&516.96	&362.38	&339.81	&341.00	&344.01	&409.39 \\
	&5	&3.65	&658.47	&506.09	&434.03	&420.35	&416.23	&411.57	&243.39 \\
	&	&	&1584.07	&832.45	&552.26	&497.58	&490.16	&485.96	&242.89 \\
	&6	&3.34	&408.37	&327.86	&285.55	&276.14	&276.70	&280.81	&377.61 \\
	&	&	&863.19	&514.41	&353.55	&316.93	&314.09	&317.52	&377.11 \\
	&8	&3.17	&475.63	&370.34	&318.53	&304.70	&303.99	&306.20	&404.97 \\
	&	&	&1114.49	&594.23	&398.22	&347.88	&338.69	&340.27	&404.47 \\ \hline
\end{tabular}
\end{center}
\label{tab:zipMLEu}
\end{table}

\begin{table}[t!]
\tabcolsep6pt
\caption{IC $ARL$ and $SDRL$ values for the ZIP-Shewhart Chart in Case U when $L$ is from Case K - MoM}
\begin{center}
\begin{tabular}{c|c|c|cccccc|c}\hline
$\phi_0$&	$\lambda_0$&	$L$	&	100	&	200	&	500	&	1000	&	2000	&	5000	&	Case K	\\ \hline
0.9	&	1	&	6.66	&	883.14	&	427.10	&	328.68	&	318.96	&	322.32	&	326.28	&	526.64	\\
	&		&		&	12359.71	&	1145.97	&	518.36	&	432.19	&	429.41	&	432.41	&	526.14	\\
	&	2	&	6.41	&	1281.62	&	624.01	&	435.30	&	398.50	&	390.07	&	389.88	&	189.92	\\
	&		&		&	13991.43	&	1953.22	&	708.40	&	527.32	&	488.35	&	486.97	&	189.42	\\
	&	4	&	5.61	&	1462.50	&	577.83	&	391.60	&	348.63	&	335.95	&	335.42	&	468.09	\\
	&		&		&	58219.25	&	1993.99	&	628.37	&	452.55	&	394.96	&	386.37	&	467.59	\\
	&	5	&	5.72	&	2839.46	&	1068.15	&	653.69	&	558.32	&	521.99	&	511.90	&	314.19	\\
	&		&		&	31349.76	&	3975.49	&	1163.99	&	762.75	&	628.38	&	590.31	&	313.69	\\
	&	6	&	5.31	&	1759.65	&	721.64	&	458.78	&	398.31	&	373.00	&	367.80	&	497.71	\\
	&		&		&	15155.06	&	2718.01	&	775.81	&	533.86	&	439.46	&	412.17	&	497.21	\\
	&	8	&	5.15	&	3049.76	&	1022.58	&	578.72	&	488.46	&	448.43	&	433.13	&	292.56	\\
	&		&		&	36613.36	&	4224.83	&	1056.43	&	674.30	&	536.22	&	481.14	&	292.06	\\ \hline
0.8	&	1	&	6.33	&	1920.78	&	1074.68	&	814.54	&	794.96	&	796.77	&	802.86	&	263.32	\\
	&		&		&	13635.01	&	3116.44	&	1285.25	&	1116.00	&	1114.25	&	1118.77	&	262.82	\\
	&	2	&	5.49	&	1647.73	&	984.21	&	727.80	&	696.39	&	694.18	&	696.20	&	301.87	\\
	&		&		&	9176.57	&	2471.15	&	1113.19	&	915.94	&	895.45	&	897.04	&	301.37	\\
	&	4	&	4.47	&	913.25	&	580.55	&	453.29	&	422.83	&	419.63	&	417.46	&	234.04	\\
	&		&		&	3924.83	&	1208.57	&	638.30	&	518.07	&	498.36	&	496.24	&	233.54	\\
	&	5	&	4.47	&	1670.69	&	948.22	&	713.63	&	652.88	&	637.28	&	636.91	&	365.09	\\
	&		&		&	9529.34	&	2154.13	&	1047.45	&	822.13	&	750.63	&	746.48	&	364.59	\\
	&	6	&	4.09	&	941.12	&	583.22	&	448.94	&	412.70	&	402.90	&	400.97	&	248.86	\\
	&		&		&	3628.02	&	1268.77	&	632.52	&	501.68	&	463.91	&	459.03	&	248.36	\\
	&	8	&	3.89	&	1157.29	&	684.89	&	509.14	&	464.15	&	445.14	&	441.07	&	289.74	\\
	&		&		&	4884.43	&	1528.66	&	730.69	&	571.11	&	506.42	&	494.43	&	289.24	\\ \hline
0.7	&	1	&	5.18	&	848.42	&	601.17	&	532.47	&	536.05	&	533.64	&	539.36	&	175.55	\\
	&		&		&	3613.84	&	1268.09	&	753.96	&	746.32	&	744.54	&	748.75	&	175.05	\\
	&	2	&	4.5	&	744.48	&	541.96	&	468.19	&	464.60	&	466.30	&	465.76	&	201.24	\\
	&		&		&	2272.43	&	1000.84	&	631.23	&	598.28	&	599.60	&	599.18	&	200.74	\\
	&	4	&	3.66	&	428.65	&	337.05	&	291.65	&	287.69	&	289.70	&	293.15	&	409.89	\\
	&		&		&	1036.65	&	549.53	&	367.38	&	339.81	&	340.25	&	343.01	&	409.39	\\
	&	5	&	3.65	&	698.85	&	514.16	&	437.42	&	419.80	&	416.28	&	410.79	&	243.39	\\
	&		&		&	1788.67	&	863.89	&	563.35	&	497.57	&	490.20	&	485.24	&	242.89	\\
	&	6	&	3.34	&	432.39	&	333.83	&	286.85	&	278.06	&	278.10	&	280.96	&	377.61	\\
	&		&		&	1015.94	&	532.35	&	357.67	&	320.08	&	315.37	&	317.64	&	377.11	\\
	&	8	&	3.17	&	509.23	&	377.29	&	322.06	&	306.22	&	303.69	&	305.87	&	404.97	\\
	&		&		&	1291.78	&	613.58	&	407.06	&	351.84	&	338.65	&	339.99	&	404.47	\\ \hline 
\end{tabular}
\end{center}
\label{tab:zipMMu}
\end{table}

The results so far show that if we want to apply the ZIP-Shewhart chart in the practice, then we must collect a very large Phase I sample, even larger than 5000 preliminary observations, in order to have similar performance between the Case K and Case U. Since this is not always possible in practice, we provide in Tables \ref{tab:zipMLEuADJ} (MLE method) and \ref{tab:zipMMuADJ} (MoM) the `adjusted' $L^{*}$ values\index{adjusted values} (see column `$L^{*}$' in the tables) which have been obtained so as the IC $ARL$ in Case U is \textit{as close as possible} to IC $ARL$ in Case K, for a given size $m$ for the Phase I sample. Note also that any other value (instead of the IC $ARL$ in Case K) can be used instead, according to practitioners needs. Once the $L^{*}$ values has been obtained, we calculate the respective IC $ARL$ and $SDRL$ values when the constant $L^{*}$ is used, instead of $L$. 

The $L^{*}$ have been determined with a two decimals accuracy, using 50000 simulation runs. The steps of the algorithmic procedure are given at Algorithm \ref{alg:algo2}. 

\begin{algorithm}
\caption{Determination of the adjusted constant $L^{*}$ - ZIP process}\label{alg:algo2}
\begin{algorithmic}[1]
\State Choose the nominal IC $ARL$ value, say $ARL_0$.
\State Collect a Phase I sample of size $m$ from a $ZIP(\phi_0,\lambda_0)$ process.
\State Use the Phase I sample and estimate $\phi_0$, $\lambda_0$ by using the MLE method (and get the $\hat{\phi}$, $\hat{\lambda}$ estimates).
\State Set the initial value $L \gets 0.01$
\State Given the values $L$, $\hat{\phi}$ and $\hat{\lambda}$, determine $\widehat{UCL}_{ZIP}$, $\widehat{LCL}_{ZIP}$ from equations (\ref{eq:UCLzip}) and (\ref{eq:LCLzip}) and calculate the IC $ARL$.
\State Repeat Steps 2-5 $T$ times and calculate the \textit{unconditional} IC $ARL$ as the average of the $T$ values $1/\left(1-\hat{\beta}\right)$.
\State Compare the $ARL$ obtained in step 6 with the $ARL_0$. If $|ARL-ARL_0|/ARL_0>0.05$ then increase $L$ by 0.01 and go back to Step 1. Otherwise, select $L$ as the `adjusted' $L^{*}$ value.
\end{algorithmic}
\end{algorithm}

For example, in the case $(\phi_0,\lambda_0)=(0.8,4)$, the IC $ARL$ ($SDRL$) in Case K is equal to $ARL = 234.04$ ($SDRL = 233.54$) with $L=4.47$. Here $ARL_0=234.04$. When the size of the Phase I sample is $m=200$, then the `adjusted' value for the chart parameter is $L^{*}=4.02$ (when the process parameters are estimated via the maximum likelihood estimation method), which gives IC $ARL=234.34$ and $SDRL=390.11$. Recall that for $m=200$ and $L=4.47$ the $ARL=566.39$ and $SDRL=1116.81$. Therefore, the use of the `adjusted' $L^{*}$ value results in an IC $ARL$ value is much closer to the nominal $ARL_0$ value while there is also a significant reduction in the $SDRL$ value.

\begin{table}[t!]
\tabcolsep6pt
\caption{`Adjusted' $L^{*}$ Values for the ZIP-Shewhart Chart - MLE method}
\begin{center}
\begin{tabular}{c|c|ccc|ccc|ccc} \hline
		    &       &    \multicolumn{3}{c}{$\phi_0=0.9$}                &		\multicolumn{3}{c}{$\phi_0=0.8$} &		\multicolumn{3}{c}{$\phi_0=0.7$} \\ \hline		
$\lambda_0$	&$m$	&$L^{*}$	&$ARL$	&$SDRL$	&$L^{*}$	&$ARL$	&$SDRL$	&$L^{*}$	&$ARL$	&$SDRL$ \\ \hline
1	&100	&6.54	&526.64	&1914.88	&4.86	&263.22	&650.62	&4.10	&179.75	&374.91 \\ 
	&200	&6.90	&528.91	&1390.08	&5.12	&262.39	&494.63	&4.26	&175.45	&276.87 \\
	&500	&7.33	&527.17	&947.08	    &5.30	&263.03	&358.17	&4.36	&174.57	&202.59 \\
	&1000	&7.51	&529.11	&739.50	    &5.38	&263.06	&288.63	&4.41	&175.39	&177.76 \\
	&2000	&7.64	&527.05	&581.89	    &5.43	&263.30	&264.45	&4.44	&175.54	&175.05 \\
	&5000	&7.67	&526.33	&526.29	    &5.64	&263.34	&262.94	&4.44	&175.55	&175.05 \\ \hline
2	&100	&4.98	&189.51	&526.49	    &4.53	&300.65	&806.72	&3.80	&199.34	&400.56 \\
	&200	&5.30	&189.66	&360.71	    &4.73	&304.39	&543.58	&3.92	&202.41	&309.87 \\
	&500	&5.52	&190.31	&265.89	    &4.86	&300.18	&406.96	&4.00	&201.09	&243.50 \\
	&1000	&5.61	&190.35	&227.59	    &4.92	&301.04	&344.89	&4.04	&201.66	&212.20 \\
	&2000	&5.67	&189.86	&200.66	    &4.96	&302.01	&309.55	&4.05	&201.21	&201.62 \\
	&5000	&5.74	&189.98	&189.86	    &4.98	&301.87	&301.42	&4.05	&201.24	&200.74 \\ \hline
4	&100	&5.14	&473.58	&3861.22	&3.88	&235.33	&579.53	&3.66	&409.93	&924.35 \\
	&200	&5.49	&463.95	&1431.24	&4.02	&234.34	&390.11	&3.75	&405.19	&654.51 \\
	&500	&5.74	&464.46	&753.44	    &4.12	&232.37	&302.78	&3.82	&406.40	&520.92 \\
	&1000	&5.84	&466.90	&618.94	    &4.15	&234.95	&273.28	&3.85	&408.91	&460.79 \\
	&2000	&5.89	&465.12	&547.49	    &4.17	&233.91	&246.11	&3.86	&407.91	&418.67 \\
	&5000	&5.94	&468.77	&487.91	    &4.19	&234.04	&234.12	&3.88	&409.90	&409.54 \\ \hline
5	&100	&4.73	&323.72	&1673.04	&3.92	&365.10	&993.44	&3.28	&242.37	&487.02 \\
	&200	&5.04	&314.12	&726.86	    &4.06	&364.24	&663.87	&3.36	&242.89	&364.17 \\
	&500	&5.24	&313.38	&480.98	    &4.16	&366.44	&499.38	&3.41	&241.53	&297.82 \\
	&1000	&5.32	&313.61	&400.84	    &4.20	&367.29	&442.19	&3.44	&245.08	&272.32 \\
	&2000	&5.36	&311.84	&359.94	    &4.22	&365.72	&398.56	&3.45	&243.57	&249.88 \\
	&5000	&5.40	&314.64	&327.18	    &4.24	&365.54	&368.39	&3.46	&243.46	&243.16 \\ \hline
6	&100	&4.81	&499.73	&2733.80	&3.62	&250.20	&689.91	&3.31	&381.12	&826.60 \\
	&200	&4.79	&492.48	&3018.95	&3.75	&250.10	&443.31	&3.39	&372.96	&594.64 \\
	&500	&5.12	&495.15	&1425.54	&3.83	&249.11	&326.73	&3.45	&380.46	&482.38 \\
	&1000	&5.45	&493.99	&663.90	    &3.86	&248.56	&292.40	&3.47	&378.22	&431.74 \\
	&2000	&5.50	&499.64	&597.13	    &3.88	&248.15	&267.48	&3.48	&376.75	&394.47 \\
	&5000	&5.53	&496.66	&530.07	    &3.89	&248.49	&250.36	&3.49	&377.58	&377.91 \\ \hline
8	&100	&4.33	&299.65	&1500.19	&3.47	&289.30	&796.77	&3.12	&404.77	&886.12 \\
	&200	&4.60	&293.34	&777.60	    &3.60	&293.82	&548.37	&3.20	&409.21	&670.61 \\
	&500	&4.80	&292.26	&463.56	    &3.68	&289.34	&390.57	&3.25	&407.76	&519.92 \\
	&1000	&4.87	&292.58	&378.29	    &3.71	&289.76	&344.97	&3.27	&409.30	&473.06 \\
	&2000	&4.91	&293.21	&340.75	    &3.73	&292.57	&323.01	&3.28	&406.63	&433.95 \\
	&5000	&4.94	&294.13	&313.23	    &3.74	&289.99	&295.78	&3.29	&405.99	&408.78 \\ \hline
\end{tabular}
\end{center}
\label{tab:zipMLEuADJ}
\end{table}

\begin{table}[t!]
\tabcolsep6pt
\caption{`Adjusted' $L^{*}$ Values for the ZIP-Shewhart Chart - MoM}
\begin{center}
\begin{tabular}{c|c|ccc|ccc|ccc} \hline
		    &       &    \multicolumn{3}{c}{$\phi_0=0.9$}                &		\multicolumn{3}{c}{$\phi_0=0.8$} &		\multicolumn{3}{c}{$\phi_0=0.7$} \\ \hline
$\lambda_0$	&	$m$	&	$L^{*}$	&	$ARL$	&	$SDRL$	&	$L^{*}$	&	$ARL$	&	$SDRL$	&	$L^{*}$	&	$ARL$	&	$SDRL$	\\ \hline
1	&	100	&	6.25	&	531.07	&	3095.24	&	4.86	&	267.77	&	857.32	&	4.09	&	175.79	&	391.94	\\
	&	200	&	6.90	&	543.54	&	1649.93	&	5.11	&	263.45	&	508.96	&	4.24	&	176.52	&	287.04	\\
	&	500	&	7.33	&	529.79	&	951.57	&	5.30	&	264.72	&	369.75	&	4.36	&	175.91	&	209.19	\\
	&	1000	&	7.51	&	532.30	&	756.62	&	5.38	&	264.53	&	296.31	&	4.42	&	175.79	&	180.14	\\
	&	2000	&	7.63	&	527.60	&	592.97	&	5.44	&	263.61	&	265.94	&	4.44	&	175.55	&	175.19	\\
	&	5000	&	7.68	&	526.70	&	528.12	&	5.32	&	263.32	&	262.82	&	4.25	&	175.55	&	175.05	\\ \hline
2	&	100	&	4.98	&	193.14	&	792.41	&	4.52	&	306.59	&	855.64	&	3.80	&	201.34	&	419.99	\\
	&	200	&	5.30	&	190.68	&	384.17	&	4.72	&	306.55	&	578.28	&	3.92	&	204.84	&	325.62	\\
	&	500	&	5.53	&	192.77	&	273.03	&	4.86	&	303.33	&	422.58	&	4.00	&	202.36	&	250.62	\\
	&	1000	&	5.61	&	190.45	&	231.20	&	4.92	&	304.04	&	357.06	&	4.04	&	201.89	&	215.39	\\
	&	2000	&	5.67	&	190.04	&	202.49	&	4.96	&	302.50	&	312.96	&	4.06	&	201.38	&	202.14	\\
	&	5000	&	5.71	&	189.98	&	189.95	&	4.97	&	301.87	&	301.48	&	4.04	&	201.24	&	400.74	\\ \hline
4	&	100	&	5.12	&	468.23	&	2383.37	&	3.86	&	235.30	&	634.38	&	3.65	&	424.18	&	1115.79	\\
	&	200	&	5.48	&	470.19	&	3398.92	&	4.02	&	237.56	&	416.71	&	3.75	&	412.83	&	687.30	\\
	&	500	&	5.75	&	471.20	&	782.49	&	4.11	&	234.15	&	309.95	&	3.82	&	412.84	&	539.78	\\
	&	1000	&	5.84	&	468.29	&	631.06	&	4.15	&	234.46	&	274.23	&	3.85	&	411.61	&	470.61	\\
	&	2000	&	5.90	&	471.43	&	561.97	&	4.18	&	235.88	&	250.54	&	3.87	&	411.04	&	425.87	\\
	&	5000	&	5.94	&	468.75	&	489.51	&	4.19	&	234.09	&	234.39	&	3.88	&	409.90	&	409.62	\\ \hline
5	&	100	&	4.71	&	320.47	&	1739.25	&	3.90	&	371.43	&	1109.44	&	3.26	&	237.90	&	480.98	\\
	&	200	&	5.02	&	314.95	&	770.66	&	4.06	&	373.64	&	710.68	&	3.36	&	248.60	&	381.95	\\
	&	500	&	5.25	&	320.30	&	500.50	&	4.16	&	366.83	&	504.38	&	3.42	&	248.55	&	309.66	\\
	&	1000	&	5.33	&	318.47	&	411.56	&	4.20	&	369.94	&	450.50	&	3.44	&	247.02	&	278.25	\\
	&	2000	&	5.37	&	316.00	&	367.03	&	4.22	&	366.28	&	401.06	&	3.45	&	244.04	&	251.96	\\
	&	5000	&	5.40	&	314.45	&	327.80	&	4.24	&	365.33	&	368.72	&	3.46	&	243.46	&	243.30	\\ \hline
6	&	100	&	4.79	&	510.34	&	4858.52	&	3.61	&	253.05	&	691.00	&	3.30	&	386.09	&	883.67	\\
	&	200	&	5.13	&	519.81	&	1693.26	&	3.74	&	249.54	&	441.93	&	3.39	&	382.38	&	621.35	\\
	&	500	&	5.37	&	509.23	&	886.18	&	3.83	&	251.35	&	333.00	&	3.45	&	384.21	&	492.48	\\
	&	1000	&	5.45	&	500.16	&	679.99	&	3.86	&	249.25	&	294.17	&	3.47	&	380.40	&	438.07	\\
	&	2000	&	5.50	&	500.60	&	601.97	&	3.88	&	250.16	&	272.42	&	3.49	&	383.03	&	405.29	\\
	&	5000	&	5.54	&	502.74	&	539.86	&	3.90	&	249.53	&	252.24	&	3.50	&	378.26	&	379.39	\\ \hline
8	&	100	&	4.29	&	299.06	&	5744.17	&	3.46	&	293.92	&	923.36	&	3.10	&	397.00	&	952.52	\\
	&	200	&	4.60	&	296.97	&	784.66	&	3.60	&	297.99	&	576.82	&	3.18	&	392.39	&	645.46	\\
	&	500	&	4.80	&	294.37	&	468.89	&	3.68	&	293.14	&	400.61	&	3.24	&	400.23	&	513.15	\\
	&	1000	&	4.87	&	293.66	&	381.99	&	3.71	&	291.57	&	349.50	&	3.26	&	400.55	&	463.44	\\
	&	2000	&	4.91	&	295.21	&	344.15	&	3.73	&	292.42	&	323.03	&	3.28	&	407.42	&	436.28	\\
	&	5000	&	4.94	&	295.41	&	315.50	&	3.74	&	290.44	&	296.90	&	3.29	&	405.94	&	408.91	\\ \hline
\end{tabular}
\end{center}
\label{tab:zipMMuADJ}
\end{table}

Before closing this section it is worth mentioning that Tables \ref{tab:zipMLEu}-\ref{tab:zipMMuADJ} offer a head-to-head comparison between the MLEs and the MEs. Recall that for the MEs there are closed-form expressions while MLEs require the use of a non-linear equation solver and can only be obtained numerically. The results show that for Phase I samples of size $m \geq 500$, the two methods provide (almost) identical results. When $m=100$ or 200, the respective $ARL$ and $SDRL$ values can be very different. According to \citet{beckett2014zero,wagh2018zero}, the MLEs are superior than the estimates obtained by the MoM, in terms of mean square error and bias, especially for small sample sizes. Therefore, we would suggest the use of MLEs when very large Phase I samples are not available. However, for larger Phase I samples (e.g. when $m>1000$) the MoM estimators of $\phi_0$, $\lambda_0$ might be more appealing to practitioners due to their simplicity in calculation.

\subsection{The Unconditional Performance of the ZIB-Shewhart Chart}
\label{sec:numericsZIB}

Tables \ref{tab:zibMLEu}-\ref{tab:zibMMu} and Tables \ref{tab:zibMLEuADJ}-\ref{tab:zibMMuADJ} are counterparts to Tables \ref{tab:zipMLEu}-\ref{tab:zipMMu} and Tables \ref{tab:zipMLEuADJ}-\ref{tab:zipMMuADJ} but for the ZIB-Shewhart chart\index{Shewhart control chart}. In Tables \ref{tab:zibMLEu} and \ref{tab:zibMMu} we present the IC $ARL$ and $SDRL$ values for the ZIB-Shewhart chart, for $n \in \left\{100,250\right\}$, $p_{0} \in \left\{0.01,0.02,0.030\right\}$, $m \in \left\{100,200,500,1000,2000,5000\right\}$ and $\phi \in \{0.9,0.8,0.7\}$. The value of $L$ has been determined in order to achieve an IC $ARL$ value in Case K that it is \textit{as close as possible} to the nominal $ARL_{0}=370.4$. Note also that the results in Table \ref{tab:zibMLEu} have been obtained via the MLE method while for those in Table \ref{tab:zibMMu} we used the MoM. \textcolor{black}{Again, the values of the achieved $ARL_0$ can be very different than the nominal 370.4 value. As for the case of ZIP-Shewhart chart, this is attributed to the discrete nature of the ZIB distribution and the large value of the zero-inflation parameter $\phi_0$.} 

\textcolor{black}{For example, for $\phi_0=0.8$, $p_0=0.01$, $n=100$ and $L=6.35$, the control limits are obtained from equations (\ref{eq:UCLzib}) and (\ref{eq:LCLzib}) and they equal $UCL_{ZIB}=3$, $LCL_{ZIB}=0$. Therefore, using equation (\ref{eq:pOOCzib}), the FAR equals $1-\beta=1-F_{ZIB}(3\left|0.8,100,0.01\right.)+F_{ZIB}(0-1\left|0.8,100,0.01\right.)\approx 0.0036748+0=0.0036748$ and the $ARL_0=1/0.0036748 \approx 272.12$}.

Conclusions from Tables \ref{tab:zibMLEu} and \ref{tab:zibMMu} are similar to those from Tables \ref{tab:zipMLEu} and \ref{tab:zipMMu}. Specifically, the IC $ARL$ values can be very different between Case K and Case U. For example, if $(\phi_0,n,p_0)=(0.9,250,0.01)$ then the IC $ARL$ and $SDRL$ are 242.82 and 242.32, respectively. But, in Case U and for $m=200$, the IC $ARL$ and $SDRL$ are 828.37 and 2962.55, respectively. Clearly, these values are much larger than the respective ones in Case K, especially the IC $SDRL$. Therefore, very large Phase I samples are needed (e.g. $m\geq 1000$) while there are cases where even $m=5000$ preliminary observations might not be enough in order to have similar performance between Case K and Case U. See, for example, the cases $(\phi_0,n,p_0)=(0.8,100,0.01)$ or $(\phi_0,n,p_0)=(0.9,100,0.02)$. However, as the size $m$ increases, the IC $SDRL$ reduces \textcolor{black}{while the IC $ARL$ converges} (not monotonically) to a specific value, which is not necessarily the same with the IC $ARL$ under Case K.

\begin{table}[t!]
\tabcolsep6pt
\caption{IC $ARL$ and $SDRL$ values for the ZIB-Shewhart Chart in Case U when $L$ is from Case K - MLE method}
\begin{center}
\begin{tabular}{c|c|c|c|cccccc|c}\hline
$\phi_0$	&	$p_0$	&	$n$	&	$L$	&	100	&	200	&	500	&	1000	&	2000	&	5000	&	Case K	\\ \hline
0.9	&	0.01	&	100	&	6.68	&	905.96	&	450.09	&	340.95	&	332.53	&	334.73	&	337.79	&	544.25	\\
	&		&		&		&	5067.77	&	1205.60	&	518.39	&	446.73	&	446.30	&	448.57	&	543.75	\\
	&		&	250	&	6.38	&	1804.08	&	828.37	&	551.49	&	490.08	&	478.06	&	475.08	&	242.82	\\
	&		&		&		&	16092.61	&	2962.55	&	916.65	&	654.97	&	590.73	&	586.15	&	242.32	\\
0.8	&	0.01	&	100	&	6.35	&	1959.66	&	1106.40	&	870.21	&	854.52	&	851.31	&	858.41	&	272.12	\\
	&		&		&		&	11762.09	&	2957.93	&	1349.42	&	1197.39	&	1193.81	&	1199.00	&	271.62	\\
	&		&	250	&	5.31	&	1845.05	&	1117.18	&	836.70	&	793.15	&	788.18	&	783.97	&	364.92	\\
	&		&		&		&	9380.28	&	2617.58	&	1248.05	&	1028.49	&	1002.35	&	998.90	&	364.42	\\
0.7	&	0.01	&	100	&	5.19	&	844.34	&	617.27	&	562.32	&	565.09	&	561.93	&	558.17	&	181.42	\\
	&		&		&		&	2883.62	&	1213.13	&	799.13	&	793.95	&	791.61	&	788.79	&	180.91	\\
	&		&	250	&	4.35	&	815.67	&	604.66	&	528.87	&	520.38	&	513.14	&	508.91	&	243.28	\\
	&		&		&		&	2210.39	&	1068.48	&	700.18	&	664.77	&	657.86	&	654.25	&	242.78	\\ \hline
0.9	&	0.02	&	100	&	6.43	&	1390.59	&	658.03	&	463.27	&	419.49	&	413.94	&	413.81	&	196.73	\\
	&		&		&		&	9792.72	&	1864.65	&	748.52	&	552.77	&	522.49	&	521.12	&	196.23	\\
	&		&	250	&	5.73	&	3301.38	&	1126.98	&	695.83	&	596.43	&	555.99	&	542.99	&	329.22	\\
	&		&		&		&	76501.60	&	3833.54	&	1231.11	&	825.55	&	668.75	&	630.13	&	328.72	\\
0.8	&	0.02	&	100	&	5.51	&	1752.68	&	1059.84	&	800.56	&	773.11	&	768.20	&	766.16	&	322.92	\\
	&		&		&		&	8044.89	&	2697.97	&	1211.15	&	1018.73	&	1000.71	&	999.11	&	322.42	\\
	&		&	250	&	4.04	&	546.84	&	367.64	&	300.80	&	284.05	&	281.48	&	284.06	&	390.77	\\
	&		&		&		&	2701.38	&	675.68	&	402.04	&	338.08	&	323.91	&	325.42	&	390.27	\\
0.7	&	0.02	&	100	&	4.51	&	771.87	&	581.34	&	509.49	&	501.80	&	499.72	&	489.92	&	215.28	\\
	&		&		&		&	2145.68	&	1034.41	&	686.15	&	659.35	&	657.09	&	648.97	&	214.78	\\
	&		&	250	&	3.66	&	730.52	&	556.84	&	475.17	&	458.28	&	456.62	&	451.26	&	260.51	\\
	&		&		&		&	1834.69	&	948.22	&	613.03	&	544.71	&	540.56	&	535.75	&	260.01	\\ \hline
0.9	&	0.03	&	100	&	6.38	&	3195.35	&	1230.92	&	756.34	&	657.13	&	624.07	&	620.12	&	320.23	\\
	&		&		&		&	55772.52	&	4905.18	&	1345.95	&	909.57	&	774.01	&	759.32	&	319.73	\\
	&		&	250	&	5.09	&	2297.93	&	808.28	&	484.48	&	414.70	&	388.05	&	375.15	&	248.86	\\
	&		&		&		&	34034.94	&	2666.92	&	834.65	&	557.19	&	460.84	&	417.27	&	248.36	\\
0.8	&	0.03	&	100	&	4.51	&	570.29	&	398.58	&	330.08	&	319.25	&	322.23	&	326.26	&	470.64	\\
	&		&		&		&	1891.84	&	746.76	&	445.74	&	390.33	&	389.40	&	392.55	&	470.14	\\
	&		&	250	&	3.86	&	935.83	&	576.64	&	439.76	&	405.54	&	391.48	&	387.97	&	252.12	\\
	&		&		&		&	3658.73	&	1244.32	&	619.49	&	494.25	&	446.75	&	439.04	&	251.61	\\
0.7	&	0.03	&	100	&	4.27	&	1083.34	&	802.81	&	680.58	&	667.35	&	668.83	&	668.83	&	313.76	\\
	&		&		&		&	3109.93	&	1467.40	&	910.28	&	842.23	&	841.27	&	840.69	&	313.26	\\
	&		&	250	&	3.4	&	1035.11	&	760.02	&	621.56	&	589.34	&	578.33	&	567.52	&	363.24	\\
	&		&		&		&	2753.87	&	1373.60	&	816.73	&	697.60	&	667.16	&	656.36	&	362.74	\\ \hline
\end{tabular}
\end{center}
\label{tab:zibMLEu}
\end{table}

\begin{table}[t!]
\tabcolsep6pt
\caption{IC $ARL$ and $SDRL$ values for the ZIB-Shewhart Chart in Case U when $L$ is from Case K - MoM}
\begin{center}
\begin{tabular}{c|c|c|c|cccccc|c}\hline
$\phi_0$	&	$p_0$	&	$n$	&	$L$	&	100	&	200	&	500	&	1000	&	2000	&	5000	&	Case K	\\ \hline
0.9	&	0.01	&	100	&	6.68	&	819.19	&	442.89	&	337.97	&	329.77	&	330.31	&	335.92	&	544.25	\\
	&		&		&		&	5757.19	&	1528.21	&	537.91	&	449.36	&	442.94	&	447.19	&	543.75	\\
	&		&	250	&	6.38	&	1964.63	&	813.32	&	553.84	&	490.51	&	476.76	&	475.55	&	242.82	\\
	&		&		&		&	38479.06	&	2549.29	&	952.35	&	663.62	&	590.62	&	586.55	&	242.32	\\
0.8	&	0.01	&	100	&	6.35	&	2066.18	&	1128.33	&	856.90	&	844.78	&	849.41	&	853.56	&	272.12	\\
	&		&		&		&	14747.90	&	3252.57	&	1376.88	&	1196.22	&	1192.42	&	1195.46	&	271.62	\\
	&		&	250	&	6	&	1907.18	&	1126.10	&	839.75	&	789.24	&	783.95	&	779.37	&	364.92	\\
	&		&		&		&	11480.70	&	2864.85	&	1286.38	&	1029.17	&	998.91	&	995.10	&	364.42	\\
0.7	&	0.01	&	100	&	5.19	&	859.75	&	620.89	&	557.01	&	555.93	&	557.25	&	553.53	&	181.42	\\
	&		&		&		&	3177.49	&	1270.84	&	798.42	&	787.10	&	788.10	&	785.27	&	180.91	\\
	&		&	250	&	4.35	&	814.89	&	609.44	&	525.10	&	517.14	&	512.98	&	503.76	&	243.28	\\
	&		&		&		&	2371.18	&	1135.87	&	710.05	&	662.48	&	657.73	&	649.79	&	242.78	\\ \hline
0.9	&	0.02	&	100	&	6.43	&	1411.89	&	664.56	&	460.10	&	417.44	&	411.19	&	410.55	&	196.73	\\
	&		&		&		&	19130.12	&	2215.13	&	772.55	&	558.28	&	520.09	&	518.42	&	196.23	\\
	&		&	250	&	5.73	&	3139.49	&	1149.55	&	691.13	&	594.82	&	554.54	&	542.81	&	329.22	\\
	&		&		&		&	36337.64	&	4337.88	&	1249.93	&	826.06	&	673.75	&	629.65	&	328.72	\\
0.8	&	0.02	&	100	&	5.51	&	1805.57	&	1050.28	&	804.90	&	763.27	&	764.15	&	770.48	&	322.92	\\
	&		&		&		&	8924.58	&	2581.10	&	1249.33	&	1010.73	&	997.52	&	1002.50	&	322.42	\\
	&		&	250	&	4.04	&	524.27	&	364.56	&	301.95	&	283.03	&	281.73	&	284.55	&	390.77	\\
	&		&		&		&	1687.15	&	693.27	&	412.39	&	337.16	&	324.26	&	325.83	&	390.27	\\
0.7	&	0.02	&	100	&	4.51	&	781.04	&	586.60	&	505.38	&	500.09	&	498.89	&	490.52	&	215.28	\\
	&		&		&		&	2441.09	&	1146.70	&	684.72	&	657.40	&	656.41	&	649.48	&	214.78	\\
	&		&	250	&	3.66	&	728.28	&	556.02	&	473.93	&	457.86	&	455.61	&	451.69	&	260.51	\\
	&		&		&		&	1916.02	&	959.71	&	618.42	&	546.08	&	539.76	&	536.13	&	260.01	\\ \hline
0.9	&	0.03	&	100	&	6.38	&	3265.02	&	1217.89	&	749.80	&	653.31	&	624.05	&	618.16	&	320.23	\\
	&		&		&		&	47087.82	&	4698.07	&	1359.02	&	923.07	&	776.75	&	757.65	&	319.73	\\
	&		&	250	&	5.09	&	2174.34	&	796.58	&	483.49	&	416.57	&	386.64	&	373.77	&	248.86	\\
	&		&		&		&	22699.92	&	3104.88	&	854.60	&	570.90	&	460.04	&	416.15	&	248.36	\\
0.8	&	0.03	&	100	&	4.51	&	571.83	&	398.43	&	329.15	&	318.58	&	319.97	&	324.40	&	470.64	\\
	&		&		&		&	2249.32	&	805.98	&	454.06	&	392.87	&	387.60	&	391.10	&	470.14	\\
	&		&	250	&	3.86	&	904.74	&	574.09	&	441.43	&	404.73	&	391.78	&	387.30	&	252.12	\\
	&		&		&		&	3503.86	&	1234.12	&	632.01	&	495.61	&	446.67	&	483.42	&	251.61	\\
0.7	&	0.03	&	100	&	4.27	&	1111.46	&	805.86	&	680.79	&	668.27	&	666.85	&	668.61	&	313.76	\\
	&		&		&		&	3585.85	&	1501.58	&	922.51	&	845.54	&	839.90	&	841.10	&	313.26	\\
	&		&	250	&	3.4	&	1029.06	&	754.55	&	622.78	&	586.09	&	578.36	&	567.83	&	363.24	\\
	&		&		&		&	2848.15	&	1366.98	&	825.29	&	695.60	&	667.58	&	656.65	&	362.74	\\ \hline
\end{tabular}
\end{center}
\label{tab:zibMMu}
\end{table}

In Tables \ref{tab:zibMLEuADJ} and \ref{tab:zibMMuADJ} we provide the `adjusted' values $L^{*}$\index{adjusted values} of the chart's design parameter $L$, so as for a pre-specified size $m$ of the Phase I sample, the IC $ARL$ under Case K is the closest possible to the IC $ARL$ under Case K. The $L^{*}$ has been determined with a two decimals accuracy, using Monte Carlo simulation. The steps of the related algorithmic procedure are similar to those presented in Section \ref{sec:CaseUZIP}, for the case of the ZIP-Shewhart chart. Some direct but necessary modifications are needed. The results in both Tables \ref{tab:zibMLEuADJ} and \ref{tab:zibMMuADJ} show that the use of the `adjusted' $L^{*}$ results in IC $ARL$ values much closer to the respective value in Case K while the IC $SDRL$ is much lower, compared to the case where the unadjusted $L$ value is used.

\begin{table}[t!]
\tabcolsep6pt
\caption{`Adjusted' $L^{*}$ Values for the ZIB-Shewhart Chart - MLE method}
\begin{center}
\begin{tabular}{c|c|c|ccc|ccc|ccc} \hline
    &   &   & \multicolumn{3}{c|}{$\phi_0=0.9$}        &  \multicolumn{3}{c|}{$\phi_0=0.8$}   &  \multicolumn{3}{c}{$\phi_0=0.7$}   \\ \hline
$p_0$	&	$n$	&	$m$	&	$L^{*}$	&	$ARL$	&	$SDRL$	&	$L^{*}$	&	$ARL$	&	$SDRL$	&	$L^{*}$	&	$ARL$	&	$SDRL$	\\ \hline
0.01	&	100	&	100	&	6.22	&	547.62	&	2898.06	&	4.83	&	275.82	&	734.92	&	4.10	&	187.96	&	416.23	\\
	&		&	200	&	6.86	&	553.26	&	1464.83	&	5.13	&	277.34	&	536.52	&	4.27	&	182.05	&	293.11	\\
	&		&	500	&	7.36	&	554.95	&	984.83	&	5.30	&	272.52	&	379.78	&	4.37	&	181.63	&	215.06	\\
	&		&	1000	&	7.53	&	550.44	&	783.88	&	5.40	&	272.71	&	300.60	&	4.43	&	181.48	&	184.54	\\
	&		&	2000	&	7.66	&	544.80	&	603.72	&	5.46	&	272.36	&	274.22	&	4.44	&	181.43	&	181.08	\\
	&		&	5000	&	7.73	&	544.28	&	544.97	&	5.34	&	272.12	&	271.62	&	4.21	&	181.42	&	180.91	\\ \hline
0.02	&	100	&	100	&	4.99	&	196.97	&	718.36	&	4.55	&	326.95	&	914.88	&	3.83	&	218.21	&	455.28	\\
	&		&	200	&	5.30	&	197.17	&	380.91	&	4.74	&	324.82	&	606.08	&	3.94	&	216.92	&	338.19	\\
	&		&	500	&	5.54	&	197.61	&	277.31	&	4.88	&	324.36	&	450.06	&	4.03	&	217.99	&	267.40	\\
	&		&	1000	&	5.63	&	196.87	&	237.15	&	4.94	&	322.97	&	373.78	&	4.06	&	216.04	&	228.13	\\
	&		&	2000	&	5.70	&	197.22	&	208.97	&	4.98	&	323.02	&	331.20	&	4.08	&	215.36	&	215.65	\\
	&		&	5000	&	5.74	&	196.78	&	196.75	&	5.00	&	322.93	&	322.55	&	4.01	&	215.28	&	214.78	\\ \hline
0.03	&	100	&	100	&	5.09	&	323.34	&	1293.61	&	4.42	&	477.90	&	1513.62	&	3.73	&	317.90	&	673.94	\\
	&		&	200	&	5.44	&	323.53	&	747.21	&	4.60	&	473.39	&	965.78	&	3.83	&	315.70	&	503.62	\\
	&		&	500	&	5.68	&	320.37	&	483.84	&	4.73	&	472.69	&	677.53	&	3.90	&	314.05	&	400.09	\\
	&		&	1000	&	5.77	&	320.28	&	415.16	&	4.78	&	471.26	&	576.32	&	3.94	&	316.69	&	349.47	\\
	&		&	2000	&	5.83	&	320.38	&	362.67	&	4.82	&	472.82	&	504.36	&	3.95	&	313.84	&	317.76	\\
	&		&	5000	&	5.88	&	320.43	&	324.91	&	4.85	&	470.93	&	471.65	&	3.96	&	313.76	&	313.26	\\ \hline
0.01	&	250	&	100	&	5.05	&	248.87	&	943.69	&	4.47	&	375.30	&	1005.00	&	3.76	&	246.02	&	511.11	\\
	&		&	200	&	5.38	&	245.81	&	525.73	&	4.65	&	368.71	&	689.54	&	3.86	&	244.62	&	379.11	\\
	&		&	500	&	5.62	&	245.20	&	356.34	&	4.79	&	370.87	&	515.43	&	3.94	&	244.92	&	304.14	\\
	&		&	1000	&	5.70	&	243.80	&	304.10	&	4.83	&	365.06	&	432.10	&	3.97	&	243.45	&	260.65	\\
	&		&	2000	&	5.76	&	243.18	&	265.94	&	4.87	&	365.57	&	381.55	&	3.99	&	243.39	&	244.93	\\
	&		&	5000	&	5.81	&	243.10	&	244.36	&	4.90	&	364.99	&	364.83	&	3.99	&	243.28	&	242.78	\\ \hline
0.02	&	250	&	100	&	4.74	&	337.48	&	1522.58	&	3.92	&	396.38	&	1240.43	&	3.29	&	262.28	&	539.24	\\ 
	&		&	200	&	5.04	&	332.37	&	875.33	&	4.07	&	393.34	&	748.01	&	3.37	&	262.76	&	397.89	\\
	&		&	500	&	5.25	&	331.23	&	511.43	&	4.17	&	393.15	&	537.20	&	3.43	&	266.14	&	331.43	\\
	&		&	1000	&	5.33	&	330.02	&	426.54	&	4.21	&	395.09	&	480.39	&	3.45	&	263.90	&	295.47	\\
	&		&	2000	&	5.38	&	331.36	&	385.70	&	4.23	&	392.28	&	427.76	&	3.46	&	261.44	&	268.93	\\
	&		&	5000	&	5.41	&	329.94	&	343.60	&	4.25	&	390.37	&	394.67	&	3.48	&	260.72	&	260.59	\\ \hline
0.03	&	250	&	100	&	4.28	&	252.68	&	1350.49	&	3.44	&	252.94	&	674.34	&	3.10	&	359.07	&	821.73	\\ 
	&		&	200	&	4.54	&	250.29	&	646.02	&	3.57	&	260.90	&	480.26	&	3.17	&	351.75	&	562.90	\\
	&		&	500	&	4.73	&	250.57	&	390.05	&	3.65	&	256.79	&	344.24	&	3.23	&	362.81	&	461.89	\\
	&		&	1000	&	4.80	&	251.69	&	321.47	&	3.68	&	257.94	&	307.20	&	3.24	&	354.55	&	406.47	\\
	&		&	2000	&	4.84	&	251.71	&	290.24	&	3.69	&	252.70	&	276.63	&	3.26	&	362.99	&	384.36	\\
	&		&	5000	&	4.87	&	252.18	&	267.24	&	3.71	&	253.99	&	258.71	&	3.26	&	362.34	&	363.33	\\ \hline
\end{tabular}
\end{center}
\label{tab:zibMLEuADJ}
\end{table}

\begin{table}[t!]
\tabcolsep6pt
\caption{`Adjusted' $L^{*}$ Values for the ZIB-Shewhart Chart - MoM}
\begin{center}
\begin{tabular}{c|c|c|ccc|ccc|ccc} \hline
    &   &   & \multicolumn{3}{c|}{$\phi_0=0.9$}        &  \multicolumn{3}{c|}{$\phi_0=0.8$}   &  \multicolumn{3}{c}{$\phi_0=0.7$}   \\ \hline
$p_0$	&	$n$	&	$m$	&	$L^{*}$	&	$ARL$	&	$SDRL$	&	$L^{*}$	&	$ARL$	&	$SDRL$	&	$L^{*}$	&	$ARL$	&	$SDRL$	\\ \hline
0.01	&	100	&	100	&	6.22	&	549.55	&	8769.46	&	4.88	&	272.38	&	892.13	&	4.12	&	182.62	&	413.66	\\
	&		&	200	&	6.92	&	544.46	&	1711.07	&	5.12	&	272.35	&	542.85	&	4.26	&	182.66	&	307.14	\\
	&		&	500	&	7.35	&	544.48	&	970.99	&	5.31	&	273.55	&	386.41	&	4.38	&	182.00	&	218.65	\\
	&		&	1000	&	7.53	&	549.34	&	794.59	&	5.40	&	273.23	&	305.94	&	4.43	&	181.43	&	185.40	\\
	&		&	2000	&	7.66	&	545.77	&	613.90	&	5.46	&	272.41	&	274.95	&	4.45	&	181.46	&	181.25	\\
	&		&	5000	&	7.73	&	544.25	&	545.20	&	5.37	&	272.12	&	271.62	&	4.26	&	181.42	&	180.91	\\ \hline
0.02	&	100	&	100	&	5.01	&	198.13	&	728.96	&	4.54	&	325.00	&	979.55	&	3.83	&	215.85	&	476.67	\\
	&		&	200	&	5.33	&	196.99	&	395.82	&	4.75	&	325.78	&	619.55	&	3.94	&	216.50	&	345.00	\\
	&		&	500	&	5.55	&	199.01	&	284.76	&	4.88	&	325.51	&	461.30	&	4.02	&	215.86	&	268.80	\\
	&		&	1000	&	5.63	&	197.01	&	240.24	&	4.94	&	324.76	&	382.56	&	4.06	&	216.52	&	231.72	\\
	&		&	2000	&	5.70	&	197.76	&	211.85	&	4.97	&	323.00	&	333.42	&	4.08	&	215.41	&	216.14	\\
	&		&	5000	&	5.73	&	196.74	&	196.76	&	5.00	&	322.95	&	322.62	&	4.01	&	215.28	&	214.78	\\ \hline
0.03	&	100	&	100	&	5.10	&	323.72	&	1962.17	&	4.42	&	477.24	&	1577.82	&	3.73	&	320.25	&	747.63	\\
	&		&	200	&	5.45	&	323.21	&	829.87	&	4.60	&	472.48	&	1011.32	&	3.83	&	315.39	&	514.54	\\
	&		&	500	&	5.69	&	320.64	&	497.95	&	4.73	&	474.02	&	692.41	&	3.90	&	314.86	&	408.38	\\
	&		&	1000	&	5.77	&	320.75	&	420.67	&	4.78	&	471.66	&	584.45	&	3.94	&	317.75	&	356.21	\\
	&		&	2000	&	5.83	&	320.69	&	366.18	&	4.82	&	473.45	&	510.22	&	3.96	&	314.94	&	321.24	\\
	&		&	5000	&	5.87	&	320.24	&	325.46	&	4.83	&	470.68	&	471.69	&	3.96	&	313.78	&	313.31	\\ \hline
0.01	&	250	&	100	&	5.07	&	244.13	&	891.46	&	4.46	&	364.97	&	1082.59	&	3.76	&	244.34	&	555.38	\\ 
	&		&	200	&	5.40	&	248.42	&	543.63	&	4.66	&	368.34	&	699.62	&	3.87	&	248.16	&	398.63	\\
	&		&	500	&	5.62	&	244.77	&	360.81	&	4.78	&	366.04	&	516.36	&	3.94	&	244.81	&	308.94	\\
	&		&	1000	&	5.70	&	243.84	&	307.25	&	4.84	&	367.57	&	442.50	&	3.97	&	243.55	&	264.50	\\
	&		&	2000	&	5.76	&	243.47	&	268.96	&	4.87	&	366.00	&	385.89	&	3.99	&	243.30	&	245.08	\\
	&		&	5000	&	5.81	&	243.20	&	244.94	&	4.90	&	365.03	&	365.05	&	3.96	&	243.28	&	242.78	\\ \hline
0.02	&	250	&	100	&	4.75	&	344.06	&	1986.21	&	3.93	&	402.80	&	1373.22	&	3.29	&	262.04	&	565.00	\\
	&		&	200	&	5.04	&	332.03	&	840.28	&	4.07	&	392.60	&	756.94	&	3.37	&	262.55	&	410.71	\\
	&		&	500	&	5.26	&	332.81	&	522.39	&	4.17	&	393.90	&	550.32	&	3.43	&	265.68	&	333.49	\\
	&		&	1000	&	5.34	&	332.74	&	432.10	&	4.21	&	395.62	&	482.49	&	3.45	&	264.35	&	298.61	\\
	&		&	2000	&	5.38	&	330.73	&	386.37	&	4.23	&	392.27	&	430.20	&	3.46	&	261.68	&	270.50	\\
	&		&	5000	&	5.41	&	330.15	&	345.08	&	4.25	&	391.27	&	394.91	&	3.47	&	260.55	&	260.38	\\ \hline
0.03	&	250	&	100	&	4.30	&	259.09	&	1453.56	&	3.45	&	258.11	&	842.59	&	3.10	&	358.13	&	868.61	\\ 
	&		&	200	&	4.55	&	250.38	&	628.92	&	3.57	&	258.28	&	484.50	&	3.18	&	362.44	&	592.21	\\
	&		&	500	&	4.73	&	249.27	&	389.73	&	3.65	&	256.53	&	345.74	&	3.22	&	351.99	&	449.17	\\
	&		&	1000	&	4.80	&	251.08	&	321.89	&	3.68	&	257.24	&	307.31	&	3.24	&	354.47	&	407.83	\\
	&		&	2000	&	4.84	&	251.44	&	290.47	&	3.69	&	253.20	&	278.18	&	3.25	&	356.52	&	377.35	\\
	&		&	5000	&	4.87	&	252.08	&	267.51	&	3.71	&	254.03	&	259.04	&	3.26	&	362.22	&	363.50	\\ \hline

\end{tabular}
\end{center}
\label{tab:zibMMuADJ}
\end{table}

\section{Out-of-Control Performance of the Charts in Case U}
\label{sec:ooc}
In this section, we investigate the OOC\index{out-of-control (OOC)} performance of the ZIP- and ZIB-Shewhart charts. It is expected that when the `adjusted' constant $L^{*}$ is used, there will be differences in the detection ability of the chart. These differences also depend on the size $m$ of the Phase I sample as well as on the magnitude of shift(s) in process parameters. Again, we use Monte Carlo simulation and the $ARL$, $SDRL$ values are provided for several shifts in process parameters. Due to space economy, only the results in the case of MLE method are presented. 

\subsection{Out-of-Control Performance of the ZIP-Charts in Case U}
\label{sec:oocZIP}
In Table \ref{tab:zipOOC}, we give the $ARL$ and $SDRL$ values for two ZIP processes: $ZIP(0.8,2)$ and $ZIP(0.7,1)$. The shifts in process parameters are $\phi_1=\tau \phi_0$ with $\tau \in \{0.8,0.6\}$ and $\lambda_1 = \delta \lambda_0$ with $\delta \in \{1.2,1.5\}$. For the first process, we assume that the size of the Phase I sample is $m=200$ while in the second $m=500$. 

The OOC $ARL$ values are compared for both ZIP-Shewhart charts\index{ZIP-Shewhart chart} with either `adjusted' or unadjusted $L$. Also, we provide the OOC $ARL$ values in Case K. The results show that the use of the `adjusted' $L^{*}$ value results in $ARL$ values that are very close to the theoretical values (i.e. those under Case K) while there is a reduction in the $SDRL$ values. Note also, that even for large shifts in process parameters (e.g. for $\delta=1.5$ and $\tau=0.6$) the respective IC $ARL$ and $SDRL$ in Case U when the unadjusted $L$ is used can be much different than the corresponding values under Case K.

\begin{table}[t!]
\tabcolsep6pt
\caption{OOC Performance of the ZIP-Shewhart Chart - MLE}
\begin{center}
\begin{tabular}{cc|cc|cc|cc} \hline
 & & \multicolumn{2}{c|}{Case K, $L=5.49$} & \multicolumn{2}{c|}{Case U, $L=5.49$, $m=200$} & \multicolumn{2}{c}{Case U, $L^{*}=4.73$, $m=200$} \\ \hline 
$\phi_1$ & $\lambda_1$ & $ARL$ & $SDRL$ & $ARL$ & $SDRL$ & $ARL$ & $SDRL$ \\ \hline
0.80 & 2.0 & 301.87 & 301.37 & 936.24 & 2109.12 & 303.87 & 549.85 \\
0.80 & 2.4 & 140.16 & 139.66 & 353.55 & 681.22  & 135.87 & 213.47 \\
0.80 & 3.0 & 59.58  & 59.08  & 121.06 & 196.63  & 55.77  & 75.56 \\
0.64 & 2.0 & 167.70 & 167.20 & 525.50 & 1226.86 & 168.79 & 301.97 \\
0.64 & 2.4 & 77.87  & 77.37  & 197.93 & 382.76  & 75.39  & 119.99 \\
0.64 & 3.0 & 33.10  & 32.60  & 67.44  & 107.50  & 31.17  & 42.15 \\
0.48 & 2.0 & 116.10 & 115.60 & 362.21 & 820.71  & 116.05 & 207.98 \\
0.48 & 2.4 & 53.91  & 53.41  & 138.30 & 283.77  & 52.26  & 81.71 \\
0.48 & 3.0 & 22.92  & 22.41  & 45.54  & 74.05   & 21.58  & 29.25 \\ \hline
 & & \multicolumn{2}{c|}{Case K, $L=5.18$} & \multicolumn{2}{c|}{Case U, $L=5.18$, $m=500$} & \multicolumn{2}{c}{Case U, $L^{*}=4.36$, $m=500$} \\ \hline 
$\phi_1$ & $\lambda_1$ & $ARL$ & $SDRL$ & $ARL$ & $SDRL$ & $ARL$ & $SDRL$ \\ \hline
0.70 & 1.0 & 175.55 & 175.05 & 537.67 & 757.06  & 174.84 & 202.52 \\
0.70 & 1.2 & 98.71  & 98.21  & 260.15 & 351.78  & 97.92  & 109.00 \\
0.70 & 1.5 & 50.78  & 50.27  & 113.41 & 145.43  & 50.17  & 53.25 \\
0.56 & 1.0 & 119.69 & 119.19 & 366.41 & 516.63  & 119.55 & 138.96 \\
0.56 & 1.2 & 67.30  & 66.80  & 177.81 & 241.10  & 66.81  & 73.99 \\
0.56 & 1.5 & 34.62  & 34.12  & 77.84  & 99.37   & 34.22  & 36.19 \\
0.42 & 1.0 & 90.80  & 90.30  & 277.36 & 391.92  & 90.46  & 104.06 \\
0.42 & 1.2 & 51.06  & 50.55  & 134.04 & 180.94  & 50.72  & 56.12 \\
0.42 & 1.5 & 26.27  & 25.76  & 59.01  & 75.39   & 25.93  & 27.24 \\ \hline

\end{tabular}
\end{center}
\label{tab:zipOOC}
\end{table}

\subsection{Out-of-Control Performance of the ZIB-Charts in Case U}
\label{sec:oocZIB}
In Table \ref{tab:zibOOC}, we provide the $ARL$ and $SDRL$ values for two ZIB processes: $ZIB(0.9,0.03,250)$ and $ZIB(0.8,0.01,100)$\index{ZIB-Shewhart chart}. Similar to the case of a ZIP process, the shifts in process parameters are $\phi_1=\tau \phi_0$ with $\tau \in \{0.8,0.6\}$ and $p_1 = \delta p_0$ with $\delta \in \{1.2,1.5\}$. For the first process, we assume that the size of the Phase I sample is $m=500$ while in the second $m=1000$. 

The results in Table \ref{tab:zibOOC} are similar to those in Table \ref{tab:zipOOC}. Specifically, they show that the use of the `adjusted' $L^{*}$ gives $ARL$ values that are very close to the theoretical ones (i.e. those under Case K), while the $SDRL$ are much lower than the respective ones in the case of `unadjusted' $L$. Like the case of ZIP processes, the OOC $ARL$ and $SDRL$ can be much larger, compared to the respective values in Case K, even for large shifts in process parameters (such as $\delta=1.5$ and $\tau=0.6$) if the `unadjusted' $L$ is used. 

\begin{table}[t!]
\tabcolsep6pt
\caption{OOC Performance of the ZIB-Shewhart Chart - MLE}
\begin{center}
\begin{tabular}{ccc|cc|cc|cc} \hline
 & & & \multicolumn{2}{c|}{Case K, $L=5.09$} & \multicolumn{2}{c|}{Case U, $L=5.09$, $m=500$} & \multicolumn{2}{c}{Case U, $L^{*}=4.73$, $m=500$} \\ \hline 
$\phi_1$ & $p_1$ & $n$ & $ARL$ & $SDRL$ & $ARL$ & $SDRL$ & $ARL$ & $SDRL$ \\ \hline
0.90 & 0.030 & 250 & 248.86 & 248.36 & 483.73 & 833.67  & 250.84 & 382.82 \\
0.90 & 0.036 & 250 & 83.19  & 82.69  & 132.07 & 187.96  & 80.43  & 104.01 \\
0.90 & 0.045 & 250 & 24.68  & 29.17  & 38.55  & 44.62   & 28.41  & 31.08 \\
0.72 & 0.030 & 250 & 88.88  & 88.38  & 173.84 & 306.16  & 89.08  & 135.42 \\
0.72 & 0.036 & 250 & 29.71  & 29.21  & 47.17  & 67.06   & 28.62  & 36.72 \\
0.72 & 0.045 & 250 & 10.60  & 10.09  & 13.78  & 15.67   & 10.12  & 10.76 \\
0.54 & 0.030 & 250 & 54.10  & 53.60  & 105.20 & 180.56  & 54.52  & 83.48 \\
0.54 & 0.036 & 250 & 18.08  & 17.58  & 28.51  & 39.97   & 17.46  & 22.24 \\
0.54 & 0.045 & 250 & 6.45   & 5.93   & 8.36   & 9.35    & 6.17   & 6.38 \\ \hline
 & & & \multicolumn{2}{c|}{Case K, $L=6.35$} & \multicolumn{2}{c|}{Case U, $L=6.35$, $m=1000$} & \multicolumn{2}{c}{Case U, $L^{*}=5.40$, $m=1000$} \\ \hline 
$\phi_1$ & $p_1$ & $n$ & $ARL$ & $SDRL$ & $ARL$ & $SDRL$ & $ARL$ & $SDRL$ \\ \hline
0.80 & 0.010 & 100 & 272.12 & 271.62 & 848.45 & 1195.33 & 272.55 & 299.14 \\
0.80 & 0.012 & 100 & 152.31 & 151.81 & 410.53 & 556.14  & 153.99 & 161.53 \\
0.80 & 0.015 & 100 & 77.86  & 77.36  & 178.83 & 229.69  & 77.64  & 80.66 \\
0.64 & 0.010 & 100 & 151.18 & 150.68 & 472.46 & 662.73  & 151.42 & 165.80 \\
0.64 & 0.012 & 100 & 84.62  & 84.12  & 229.06 & 309.98  & 89.65  & 90.28 \\
0.64 & 0.015 & 100 & 43.26  & 42.75  & 98.62  & 127.01  & 43.08  & 44.46 \\
0.48 & 0.010 & 100 & 104.66 & 104.16 & 325.56 & 458.49  & 105.11 & 115.89 \\
0.48 & 0.012 & 100 & 58.58  & 58.08  & 158.08 & 213.82  & 58.63  & 62.54 \\
0.48 & 0.015 & 100 & 29.95  & 29.44  & 68.74  & 88.10   & 29.86  & 30.80 \\ \hline

\end{tabular}
\end{center}
\label{tab:zibOOC}
\end{table}

\section{Conclusions}
\label{sec:conclusions}
In this work we studied the design and performance of Shewhart-type charts for zero-inflated processes, when the process parameters are unknown and must be estimated. Using Monte Carlo simulation and two estimation methods (maximum likelihood and method of moments) we evaluated the performance of the considered charts for several Phase I sample sizes, in terms of the unconditional $ARL$ and $SDRL$ performance measures. The numerical results showed that in the case of zero-inflated processes, very large Phase I samples are needed in order to obtain a reliable design of the chart, whose performance in the case of the estimated parameters is similar to its theoretical performance. In almost all of the examined cases, the suggested size $m$ of the Phase I sample is larger than 1000 preliminary observations while there are cases where even more than 5000 preliminary observations do not guarantee that the actual performance will be close to the theoretical one. 

For relatively small preliminary samples (e.g. for $m=100$ or 200), the IC $ARL$ and $SDRL$ values can be very large. Therefore, we do not advise practitioners to set-up their chart when the process parameters are estimated from a Phase I sample of this size. Note also that in case of large $\phi_0$ values, a small Phase I sample might end up to a sample with only zero values, making the estimate of process parameters impossible. As a general guideline, we suggest that Phase I samples with at least 1000 observations are necessary to obtain reliable estimates of process parameters. As $m$ increases, the IC $ARL$ and $SDRL$ reduce, but not necessarily in a monotonic way (the former). 

\textcolor{black}{In an attempt to assist practitioners in ending up with reliable designs for their charts}, we provided `adjusted' values \textcolor{black}{for the design parameter} $L$ in order to obtain an IC $ARL$ very close to IC $ARL$ in Case K, for a given size $m$ of the Phase I sample. Note also that the proposed methodology works for any other nominal $ARL_0$ (instead of the IC $ARL$ value in Case K) that meets \textcolor{black}{practitioners'} needs. 

Finally, We used R (see \citet{rproject2023}) for all the necessary computations in this work. The related programs are available from the authors upon request.



\end{document}